# GenAI Distortion: The Effect of GenAI Fluency and Positive Affect

Xiantong Yang [1,] Mengmeng Zhang [2, *]

[1] Faculty of Psychology, Beijing Normal University, Beijing 100875, China

[2] Faculty of Education, Minzu University of China, Beijing 100081, China

**Contact Information for all authors:**

Xiantong Yang: xtyang93@foxmail.com

Mengmeng Zhang: zhangmm817@foxmail.com (Corresponding Author)

## Abstract

The introduction of generative artificial intelligence (GenAI) into educational practices has been transformative, yet it brings a crucial concern about the potential distortion of users' beliefs. Given the prevalence of GenAI among college students, examining the psychological mechanisms that lead to GenAI distortion from both the technological factors and the individual's psychological processes is a critical priority. A mixed-methods approach is employed to test the proposed hypotheses. Study 1 ($N = 10$) revealed through qualitative analysis that GenAI's fluent outputs significantly engaged college students, eliciting positive emotional responses during an interaction. GenAI's tendency to conflate fact with fiction often led to presentations of unrealistic and exaggerated information, potentially distorting users' perception of reality—a phenomenon termed GenAI distortion. Following these insights, Study 2 (cross-sectional survey, $N = 999$) and Study 3 (experimental manipulation, $N = 175$) explored how GenAI fluency affects college students' GenAI distortion and examined the mediating effect of positive affect. The results indicated that GenAI fluency predicts GenAI distortion via the mediating role of positive affect. Our findings provide theoretical foundations and practical implications for understanding GenAI distortion among college students.

## 1. Introduction

Generative artificial intelligence (GenAI), encompassing technologies like OpenAI's GPT series, Google's Bard, Stable Diffusion, and Midjourney, has captured the minds of the public and inspired widespread adoption (Kidd & Birhane, 2023). GenAI has fundamentally changed the landscape of educational practices, experiencing continuous proliferation in education (Crompton & Burke, 2023) owing to their potential benefits, including improved personalized learning (Bhutoria, 2022), provision of efficient administrative processes (Parycek et al., 2023), assist computer programming and text writing, etc. (Sun et al., 2024).

However, such advancement comes with some concerns, as scientists have raised issues regarding the data sources in GenAI that may distort scientific facts, leading to potential biases and the dissemination of false information (Kidd & Birhane, 2023; Rawas, 2023). Individuals may believe the content generated by GenAI systems, leading to distortion from normal cognition or understanding. This distortion may harm users' cognitive and logical skills and even lead to a decline in academic competency (İpek et al., 2023).

Cognitive distortions, labeled as dysfunctional thinking patterns or beliefs (Zhang, 2008), are irrational and inaccurate perceptions of oneself or one's environment (Briere, 2000), commonly applied in the field of mental health. In the context of Gen AI, the research scope on cognitive distortions has broadened to encompass usage scenarios of technologies like GenAI (Extance, 2023; Kidd & Birhane, 2023). Kidd and Birhane (2023) highlighted a concerning issue that GenAI models may distort human beliefs due to their inability to reliably distinguish between facts and fiction. Considering the characteristics of GenAI, we define GenAI distortion as the cognitive bias that individuals may experience due to the influence of GenAI algorithmic features on the content it generates, leading to deviations from normal understanding. In other words, when individuals interact with GenAI conversational systems, the training data for GenAI comes from a large body of text on the internet, which can include misleading, inaccurate, or biased information (Kidd & Birhane, 2023).

Individuals may find it difficult to identify potential errors or inaccuracies, collectively referred to as GenAI distortion. For example, in the fields of science and engineering, objectively, they are not entirely male-dominated, but GenAI often uses male pronouns when describing these professions, which is a specific manifestation of GenAI distortion. The trend of anthropomorphizing technology increases the presentation of GenAI models' unrealistic and exaggerated capabilities, leading to widespread misunderstandings and exacerbating the risk of spreading false information and negative stereotypes to people. Given the prevalence of GenAI among college students, researching how GenAI distorts students' cognitive beliefs is a significant priority. It's crucial to investigate the psychological mechanisms that lead to GenAI distortions among college students from both the perspective of the technology itself and the individual's emotional and cognitive processes.

## 2. Overview of the studies

Previous research has explored concepts related to cognitive distortion (Cacault & Grieder, 2019; Zhang, 2008), but no empirical studies have yet proposed and verified the academic concept of GenAI distortion within the context of GenAI. Moreover, research examining the factors that contribute to distortion during user interactions with GenAI is lacking, prompting our mixed study.

We employed a three-stage mixed-methods approach. In Study 1, semi-structured interviews were conducted to investigate the distortions experienced by university students using GenAI and to explore the associated psychological processes. This approach deepened comprehension of GenAI distortion concepts, uncovered relational mechanisms, and provided the groundwork for subsequent questionnaire surveys and experimental studies. Qualitative data analysis also refined measurement tools and manipulation methods for future quantitative research, enhancing studies' validation and reliability (David et al., 2018). In Study 2, we conducted a questionnaire survey among college students familiar with GenAI tools like ChatGPT. The aim was to assess the extent of GenAI distortion and to verify the associations between various variables. In Study 3, we developed online experimental tasks to further validate the results of the questionnaire survey. By combining qualitative and quantitative methods, our study delved into the psychological mechanisms behind GenAI distortion in college students. These findings would offer valuable insights for exploring effective strategies to reduce GenAI distortion in the future.

### 3. Study 1: A qualitative study

Given the scarcity of research on GenAI distortion, qualitative research offers an opportunity to thoroughly understand and describe the topic. It can uncover specific manifestations of GenAI distortion and shed light on crucial factors that have been overlooked or not previously identified. This groundwork is essential for informing subsequent quantitative studies.

### 3.1. Participants and procedure

A semi-structured interview research design has been developed. We recruited research participants through the public platform *https://www.credamo.com*, adhering to the principle of voluntary participation. All measurements and procedures were approved by the Institutional Review Board (IRB) of the first author's institution. This study used purposive sampling, and participants should have experience in using GenAI but with varying characteristics in terms of age, and educational background.

One trained and experienced researcher presided over the semi-structured interviews, scheduling meeting times based on their availability. The online interview duration (Zoom and Tencent Meeting) ranged between 30 and 60 minutes. At the beginning of the interview, the researcher explained the interview's purpose. Subsequently, the researcher posed several questions to obtain the participants' opinions on this topic. The conversations were documented and transcribed. Interviews continued until reaching saturation standards (Malterud et al., 2016). Table 1 shows the demographic information of the participants.

Table 1. Descriptive information of participants

| No. | Sex | Age | Major | Degree |
|-----|--------|-----|-------------------------|----------|
| 1 | Female | 18 | Education | Bachelor |
| 2 | Female | 20 | Social work | Bachelor |
| 3 | Female | 21 | Education | Bachelor |
| 4 | Female | 25 | Business Administration | Mater |
| 5 | Female | 24 | Linguistics | Mater |
| 6 | Male | 25 | Computer Engineering | Mater |
| 7 | Male | 27 | Education | PhD |
| 8 | Male | 30 | Psychology | PhD |
| 9 | Male | 28 | Education | PhD |
| 10 | Male | 28 | Computer Engineering | PhD |

### 3.3. Data analysis

Two researchers independently coded interview transcripts to identify core themes and illustrative statements. We utilized open coding as part of our thematic analysis, systematically examining the data line by line to identify terms and concepts expressed by college students, which were then treated as initial codes. Additionally, we categorized some related terms as main themes. If necessary, these themes were further subdivided into sub-themes, incorporating the initial coding categories. Our goal was to determine the main themes of the discussions by identifying commonalities and grouping key aspects within each theme. Following the coding of the transcripts into predefined categories, we discussed any discrepancies in our analyses until reaching a consensus.

**3.4. Results**

Table 2 summarizes the main themes from the semi-structured interviews and provides explanatory statements. The first question focuses on why college students utilize GenAI. Generally, most users regard GenAI (e.g., ChatGPT) as their study assistant because it delivers prompt feedback when they inquire. This efficiency prompts many college students to rely on GenAI for tasks like completing assignments and addressing questions.

The second set of questions evaluates participants' experiences with GenAI. There is a strong consensus that, although GenAI is user-friendly and potent, the algorithmic nature of GenAI systems may lead to some incorrect answers. If users lack discernment, GenAI distortion may emerge, manifesting in various ways.

The final question extends the second one, exploring the factors that cause users to develop distortion while using GenAI. Some college students noted that when they ask GenAI questions, it frequently delivers prompt answers, creating the impression of a dependable assistant. This smooth experience frequently elicits positive emotions in users. Once they form a positive emotional bond, they may trust GenAI's content without recognizing the potential for cognitive biases.

Table 2. Results of the interview

| Themes | Description | Example participant statements |
| --- | --- | --- |
| GenAI fluency | Naturalness: Users perceive that the text generated by GenAI, such as ChatGPT, is similar to human language expression, without seeming stiff or robotic. | I once asked "ERNIE Bot" about the process of planning activities, and its responses were well-organized, using language similar to what I use in regular conversations, making me feel comfortable (Participant 2). Dialogues with ChatGPT feel natural and I see it more as a companion than merely a program (Participant 5). |
| | Fluency: Users feel that the text generated by AI is smooth, enhancing the natural fluency of the conversation. | During finals, I was working on a paper about the internationalization process of higher education in the United States. My grasp of the topic wasn't very clear. However, when I asked ChatGPT questions related to it, it gave coherent and detailed answers, saving me a lot of time (Participant 9). While ChatGPT occasionally gave inaccurate responses, it consistently provided smooth answers whenever they posed questions (Participant 10). |
| | Fast responsiveness: Users perceive that GenAI responds quickly, without requiring long wait times, thereby enhancing user satisfaction. | When I ask a question, ChatGPT or ChatGLM quickly provides a response, which helps maintain a smooth conversation flow (Participant 1). Faced with a new academic concept, I naturally turn to GenAI for answers. ChatGPT's rapid response gives me a sense of authenticity in the information it provides (Participant 8). |
| Positive affect | Pleasure: Users feel enjoyment in interacting with GenAI due to its responses containing humor, friendliness, or interesting expressions. | Using ChatGPT, I get quick responses to my input without long waits, which gives me a very enjoyable experience (Participant 4). |
| | Satisfaction: The answers provided by GenAI are typically satisfactory. | I am highly satisfied with ChatGPT. It has provided me with some helpful suggestions, and I will continue to rely on it and recommend it to those around me (Participant 6). |
| | Ease: Conversing with ChatGPT is an effortless, stress-free experience. | In conversations with ChatGPT, I can ask any question without any hesitation or concern (Participant 3). |
| | Emotional Connection: GenAI can recognize and respond to users' affect, making the conversation more humane. | Once, feeling disheartened after criticism from my supervisor, I turned to ChatGPT for advice. Not only did it offer suggestions, but it also showed understanding and provided encouragement, as if sincerely listening to my feelings. This emotional connection brought me comfort and increased my dependence on ChatGPT (Participant 5). ChatGPT, as a quiet confidant, provides me with emotional understanding and encouragement (Participant 2). ChatGPT has become a reliable companion for me because there are topics I can't discuss even with close friends. ChatGPT doesn't judge me and offers emotional support (Participant 4). |

| The GenAI distortion manifests as diversity | Trust bias: Due to their trust in the technology, the user is inclined to accept information generated by GenAI without exercising sufficient caution in evaluating its accuracy. | Initially, I trusted ChatGPT's judgments on some widely accepted facts, so I consulted a statistical question involving β and B. Although its response seemed reasonable, I later discovered inaccuracies during class, resulting in errors in my assignments (participant 7). |
|---|---|---|
| | Misleading bias: Users ask misleading questions, and GenAI generates text based on incorrect assumptions, affecting users' understanding of the generated content. | I once asked ChatGPT a question about deep learning, but my question may not have been accurately described. ChatGPT provided a misleading answer, and I mistakenly believed it to be true (Participant 5). The question's complexity was beyond my current comprehension, which made me unaware of the potential bias in the answer (Participant 6). |
| | Unconscious bias: GenAI exhibits societal and cultural biases, but users are unaware that these biases exist. In this context, users may unintentionally accept these biases in GenAI's responses. | In my European cultural history assignment, I emphasized the achievements of white culture using information from ChatGPT. However, my teacher criticized my writing for being biased, pointing out that I focused excessively on white culture and ignored contributions from other races and cultures. Due to limited understanding and time constraints, I didn't fact-check and directly used information from ChatGPT, which resulted in criticism of my assignment (Participant 4). |

## 3.5. Discussion

This exploratory study explores how users perceive the GenAI. Currently, GenAI is extensively used, but several considerations emerge. Firstly, users face cognitive distortions when using GenAI, manifesting in various ways. Secondly, the fluency experience of GenAI enhances users' positive affect. Even if the GenAI system may have errors or biases, however, due to this positive impression, users may not be able to accurately identify issues in GenAI-generated content. Thirdly, the fluency of GenAI can evoke positive emotions in users, particularly when handling unfamiliar topics, thus potentially exacerbating cognitive distortions. Regarding the validity of these findings in large-scale surveys, we utilized quantitative research in Study 2, employing a cross-sectional research design to confirm the qualitative research findings.

## 4. Study 2: A quantitative study

## 4.1. Hypotheses development

Study 1 indicated that individuals perceived GenAI fluency and positive affect may be related to cognitive distortion. We will next comprehensively present an argument for the research hypothesis, drawing on the results of qualitative analysis, relevant literature, and theories.

### 4.1.1. GenAI fluency and GenAI distortion

In the qualitative study, we found that the fluent responses from GenAI contribute to users' increased acceptance and satisfaction of generated content, even when it may be inaccurate or misleading. Fluent information processing, while associated with quicker judgment, lacks thoughtful cognitive processing, potentially fostering distortions in comprehension.

GenAI fluency refers to the ease with which individuals perceive the external, surface, and physical features of stimuli during the perceptual stage of cognitive processing (Reber et al., 2004; Winkielman et al., 2003). The dual-process theory (Alter et al., 2007) and the assumption that disfluency triggers analytic thinking (Alter & Oppenheimer, 2006) suggested that perceptual disfluency increases individuals' tendency to overcome intuitive responses, leading to more rational reactions and then reducing cognitive distortions. Similarly, some studies indicated that higher perceptual fluency can strengthen individuals' intuitive reactions, but may also lead to misleading, biased, or inaccurate understanding (i.e., cognitive distortion) (Hansen et al., 2008; Novemsky et al., 2007). For example, Hansen et al. (2008) manipulated color contrast to control perceptual fluency. The results revealed that sentences with higher perceptual fluency were more often perceived as true compared to those with lower perceptual fluency. Specific to the context of GenAI, fluency often creates an illusion of accuracy and reliability, which can make users more inclined to accept the generated content without questioning it, thereby overlooking potential errors or biases. Therefore, we hypothesized that GenAI fluency may have a positive impact on GenAI distortion.

4.1.2. The mediating role of positive affect

In Study 1, users expressed satisfaction with GenAI's prompt, frequent, and naturally fluent language responses, demonstrating positive emotions. This suggests that the fluency of GenAI's language positively influenced users' positive experiences. Furthermore, we found that satisfied users tend to trust GenAI's responses, even with the potential for misleading, biased, or inaccurate information.

Prior studies suggested that high perceptual fluency leads to increased positive affect for the assessed stimuli (Reber et al., 1998; Albrecht & Carbon, 2014). As a result, individuals were more likely to attribute this perceptual fluency to positive features rather than negative ones (Reber et al., 1998; Albrecht & Carbon, 2014). Similarly, Topolinski and Strack (2009) also found that when stimuli are processed fluently, the individual doing the processing is inclined to exhibit a subtle smile (i.e., positive affect). In other words, GenAI fluency may be positively associated with positive affect. Furthermore, affect as information theory proposes that individuals' emotions are strong determinants of their cognitive evaluations and that they automatically integrate these states into their decision-making processes (Schwarz, 2012). GenAI distortion is a prerequisite for inaccurate individual judgments. Therefore, positive affect may be positively associated with GenAI distortion.

Integrating the aforementioned indirect evidence on the relation between GenAI fluency and positive affect, as well as the relation between GenAI fluency and GenAI distortion, the intermediary influence mechanism of positive affect becomes apparent. Based on the hedonic fluency model (Winkielman et al., 2003), the increasing perceptual fluency of a stimulus leads to more positive judgment, mediated through the positive affect (i.e., fluency-affect link). This is because the hedonic quality of perceptual fluency (Winkielman et al., 2003) itself influences one's affective state, and personal affect serves as information aids in judgment (Schwarz & Clore, 1983, 2007). Considering the GenAI distortion is a prerequisite for inaccurate individual judgments. We can hypothesize that GenAI fluency may increase friendliness and satisfaction with GenAI, and this positive affect may weaken users' perception to detect potential errors in GenAI responses, leading to biased cognition. In other words, positive affect plays a mediating role in the relationship between GenAI fluency and GenAI distortion.

## 4.2. Methods

### 4.2.1. Participants

In this study,  a total of 1062 participants completed the psychometric scales, and 999 participants (647 boys, 352 girls) successfully passed the quality check items (e.g., "Please select the first option for this item."). They aged from 18 to 34 years old (M = 22.74, SD = 3.25). Participants were recruited through an online platform (Google Forms) and completed online self-reported questionnaires voluntarily, which took approximately 10 minutes. We also provided with an informed consent form before completing the survey.

### 4.2.2. Measures

**GenAI fluency**. This study adopted a GenAI fluency questionnaire to assess students' levels of perceptual fluency when using GenAI, such as ChatGPT, and Google's Bard. The initial questionnaire had eight items (Yoo & Kim, 2014). The original items were subjected to first-order Confirmatory Factor Analysis (CFA) to evaluate their appropriateness in measuring their respective constructs. Following the assessment of the first-order model fit, the questionnaire items were scrutinized for residual values. If two items were correlated, the residuals would not be independent (Kline, 2016). In this study, three items with the same meanings within the construct were identified and removed. The final five items (e.g., "*GenAI presents information clearly*") were rated on a 5-point scale, ranging from 1 (*strongly disagree*) to 5 (*strongly agree*), to assess perceptual fluency. Higher scores indicated that individuals perceive higher levels of fluency in GenAI. The current sample revealed a good internal consistency (Cronbach's α = 0.72; Omega = 0.74).

**GenAI distortion.** The study adopted a GenAI distortion questionnaire to measure college students' distortion tendencies while using GenAI (Chung & Han, 2013). The initial questionnaire consisted of eight items, and four items that did not meet the criteria were deleted. Conduct CFA for the remaining four items (e.g., "*ChatGPT has started to possess self-awareness*"), the results indicated a good fit, $\chi^2/df$ = 1.023, $p < 0.001$, RMSEA = 0.005, CFI = 1.000, TLI = 1.000, SRMR = 0.007. Five-point Likert scales (1 = "*strongly disagree*", 5 = "*strongly agree*") were used. Because the descriptions of these four items deviate from objective reality (e.g., ChatGPT does not possess self-awareness), participants who exhibited more severe cognitive biases and greater distortions associated with general AI would score higher. The current sample revealed a good internal consistency (Cronbach's α = 0.7; Omega = 0.72). In the pilot study, this scale showed a moderate correlation ($r = 0.712$, $p < 0.001$) with previous measures of folk psychological attributions of consciousness to large language models (LLMs), which are considered a cognitive bias (Colombatto & Fleming, 2024), indicating reasonable construct validity.

**Positive affect.** This study explores the positive affect that users experience when interacting with GenAI, building on the research conducted by Kern et al. (2015). CFA revealed a good fit, $\chi^2/df$ = 4.257, $p < 0.001$, RMSEA = 0.057, CFI = 0.990, TLI = 0.970, SRMR = 0.016. Four items were used to measure the users' positive affect (e.g., "*I feel very enjoyable while using ChatGPT*").

Five-point Likert scales (1 = "*strongly disagree*", 5 = "*strongly agree*"). The present sample revealed a suitable internal consistency (Cronbach's α = 0.72; Omega = 0.73).

### 4.2.3. Data analysis

All analyses were performed using the *lavaan* package (versions 0.6-15) in R studio software (version 4.3.1). The *lavaan* package serves as a powerful tool for performing Structural Equation Modeling (SEM). It empowers researchers to estimate and fit intricate models, facilitating the exploration of association among latent variables. Firstly, descriptive and correlational analyses were conducted. Secondly, we use the *sem()* function to estimate model parameters and fit the mediation model. The *summary()* function is then employed to assess the goodness of fit for the model (e.g., $\chi^2/df$, CFI, TLI, RMSEA, and SRMR).

## 4.3. Results

### 4.3.1. Preliminary analyses

Table 3 showed significant correlations among sex, age, and main variables, indicating that subsequent analyses need to take sex and age into consideration as control variables. Regarding the main variables, GenAI fluency, GenAI distortion, and positive affect were positively correlated with each other (*r* ranged from 0.326 to 0.633).

Table 3. Descriptive characteristics and correlation (*N* = 999).

|  | *M* | *SD* | 1 | 2 | 3 | 4 | 5 |
|---|---|---|---|---|---|---|---|
| 1. Sex | 1.352 | 0.478 | 1 |  |  |  |  |
| 2. Age | 22.743 | 3.251 | 0.127* | 1 |  |  |  |
| 3. GenAI fluency | 3.902 | 0.581 | 0.1** | 0.196** | 1 |  |  |
| 4. GenAI distortion | 3.589 | 0.745 | 0.034* | 0.219** | 0.326*** | 1 |  |
| 5. Positive affect | 3.908 | 0.625 | 0.07* | 0.144** | 0.633*** | 0.364*** | 1 |

Note. *p < 0.05, **p < 0.01, ***p < 0.001.

### 4.3.2. Testing the mediation of positive affect

To examine the effects of positive affect between GenAI fluency and GenAI distortion, a mediation model was conducted. According to the suggestion of Cheung (2007), the mediation model showed a satisfactory model fit [$\chi^2/df$ = 2.839, CFI = 0.950, TLI = 0.938, RMSEA (90% CI) = 0.043 (0.037-0.049), SRMR = 0.052]. The results revealed that GenAI fluency could positively predict positive affect ($\beta$ = 0.901, $p$ < 0.001), but did not significantly predict GenAI distortion ($\beta$ = -0.034, $p$ > 0.05). Positive affect could positively predict GenAI distortion ($\beta$ = 0.542, $p$ = 0.001). Further analysis adopted a bootstrapping method with 2000 bootstrap samples, the mediated effect value of positive emotion was 0.488 ($p$ = 0.002, 95% CI = [0.258, 0.879]), statistically greater than zero, indicating that positive affect could mediate the association between GenAI fluency and GenAI distortion.

### 4.4. Discussion

Study 2 used the self-report scales to explore the connections between key variables. Our findings suggest that GenAI fluency predicts positive affect, which in turn predicts GenAI distortion. That is, positive affect mediates the link between GenAI fluency and GenAI distortion. Yet, Study 2 collected data using a cross-sectional design, which may not fully address the causal relations between variables. In the subsequent Study 3, we used experimental manipulation to examine the underlying mediating process through experimental control.

## 5. Study 3: An experimental study

In Study 3, we manipulated the independent variable (GenAI fluency) and measured the mediating variable (positive affect) and the dependent variable (GenAI distortion). Specifically, we adopted a between-subjects single-factor experimental design. We manipulated students' perception of ChatGPT fluency based on an association test by constructing two scenarios using GenAI. Our study has been preregistered on OSF (https://doi.org/10.17605/OSF.IO/2RHQN).

## 5.1. Methods

### 5.1.1. Participants

Before the experiment, we conducted a priori sample size calculation using G*power 3.1 (Faul et al., 2007). With an alpha (α) of 0.05, a power of 0.8, and five predictors, we determined that a total sample size of 92 was needed to detect a medium effect size ($f^2 = 0.15$) in linear multiple regression using a fixed model with $R^2$ deviation from zero. The initial sample included 185 college students. Following prior research recommendations (Arechar et al., 2023), we added an attention check question to identify students who did not respond attentively. Six participants didn't pass the attention check. During data cleansing, four participants were removed because of missing data. We finally included 175 students in the analyses. Their average age was 20.75 years ($SD = 2.71$), ranging from 18 to 29 years; 41 (23.4%) were females and 134 (76.6%) were males. This study has been approved by the Academic Ethics Committee of the university of the first author.

### 5.1.2 Procedure and materials

Students voluntarily participated in this experimental study conducted online via the Online data collection platform, which took approximately 20 minutes. Initially, we conducted a pre-test on all students participating in the survey to measure their objective cognitive distortion, subjective cognitive distortion, current emotional states, and demographic information containing sex, age, etc. Then, participants were presented with a scenario describing the scene when the user talks to the GenAI. One group of participants was randomly assigned to a scenario where the response speed of the GenAI was very fast and smooth, while another group of participants was randomly assigned to a scenario where the response speed of the GenAI was very slow and laggy. This imagery paradigm has been considered effective in manipulating fluency by previous research (Alter & Oppenheimer, 2009).

In the experimental group, participants read the following paragraph depicting a scenario where they imagined themselves using GenAI very fluently:

"*Picture yourself chatting with a GenAI. Here, the AI's responses are lightning-fast and seamless. You ask a question, and within seconds, the AI delivers detailed, precise answers. Interacting with the AI feels effortless, with almost no lag. You see the AI as a smart, efficient assistant that grasps and responds to your needs promptly during your exchanges.*"

In the control group, participants read the following paragraph depicting a scenario where they imagined themselves using GenAI very slowly:

"*Imagine you're using a generative AI for a conversation. In this scenario, the AI's responses are slow, often lagging. You must wait a long time for answers to your questions, and sometimes the responses are not detailed or accurate enough. The interaction with the GenAI feels very clunky, with frequent delays. In your exchanges with the GenAI, it comes across as an inefficient, slow-to-react tool that struggles to quickly understand and meet your needs.*"

After reading the scenario, all participants were asked to answer three manipulation check questions about their perceived GenAI fluency. Subsequently, participants reported their positive affect, objective GenAI distortion, and subjective GenAI distortion.

5.1.3. Measures

Pre-test variables

**Subjective GenAI distortion.** The four items for subjective GenAI distortion are the same as in Study 2. Participants were asked to evaluate their level of agreement with the following statements, according to their actual thoughts (e.g., "*ChatGPT has started to possess self-* *awareness*”), with 1 representing strongly disagree and 5 representing strongly agree, where a higher score indicates a higher subjective distortion. The Cronbach'α reliability coefficient was 0.723.

**Objective GenAI distortion.** We asked participants to evaluate the accuracy of statements about GenAI. We used ten questions to score objective distortions about GenAI, assigning 1 point for incorrect answers and 0 points for correct answers. Ten items measuring objective GenAI distortion were evenly distributed between the pretest and posttest, with five items in each. For example: “*GenAI has autonomous decision-making capabilities and can make accurate decisions without human intervention*.” The items in the pretest and posttest are balanced in difficulty, with no significant differences. We sum the error scores of the five items from both the pretest and posttest separately and then calculate the average. A higher score indicates a greater level of GenAI distortion. The scale demonstrated an internal consistency reliability of 0.771 in the pilot study, and its correlation with subjective cognitive distortions related to general AI was $r = 0.412$ ($p < 0.001$), proving that both reliability and construct validity are acceptable.

**Perceived task difficulty.** To assess the difficulty of objective cognitive distortion tasks, we followed Efklides (2006) and Urban et al. (2024) by using one item (“*How difficult was it to solve this question?*”) to ask participants to rate the difficulty of objective questions on a scale from 1 (very easy) to 5 (very difficult), where a higher score indicates a harder question.

**Current emotional state.** To mitigate the potential effect of participants' current emotional states on the study findings, we assessed and controlled for their current emotional state. Following the study of Kern et al. (2015), four questions were included in this study (e.g., “I feel very pleased right now.”). The Cronbach's α reliability coefficient was 0.917.

Post-test variables

**Manipulation check.** Participants in both groups completed three manipulation check questions to assess their perceived level of GenAI fluency. Participants responded to these three questions using a 5-point Likert scale, ranging from 1 (*strongly disagree*) to 5 (*strongly agree*). Participants in the experimental group reported a higher level of perceived GenAI fluency compared to those in the control group.

**Positive affect.** Participants were given the following instructions for the scale: "In the imagined scenario described above, please evaluate the pleasantness of using GenAI:..." In this experimental study, we included four items to assess participants' positive. The Cronbach's α reliability coefficient was 0.919.

**Subjective GenAI distortion.** The test questions were consistent with those in the pre-test. However, we made modifications to the instructions in the guide language. Participants were required to evaluate their agreement with the following statements based on the imagined scenario provided above. In the post-test, the Cronbach's α reliability coefficient was 0.856.

**Objective GenAI distortion.** This test remained consistent with those in the pre-test. However, we made modifications to the instructions in the guide language. "Based on the imagined scenario provided above, please evaluate the following statements to determine which ones are correct and which ones are incorrect."

### 5.2. Results

#### 5.2.1. Baseline test analysis

We conducted an independent samples t-test on the objective GenAI distortion and subjective GenAI distortion between the experimental and control groups in the pretest. The results showed that there were no significant differences in either subjective GenAI distortion ($M_{experimental}$ = 2.60, $M_{control}$ = 2.42, $t$ = -1.71, $p$ = 0.089) or objective GenAI distortion ($M_{experimental}$ = 0.48, $M_{control}$ =

0.28, $t$ = -1.86, $p$ = 0.065) between the experimental and control groups. This indicates that the experimental and control groups had similar states before the manipulation.

### 5.2.2. Manipulation check

As shown in Table 4, an independent-sample $t$-test revealed that participants in the experimental condition reported more ChatGPT fluency than participants in the control condition ($M_{experimental}$ = 3.26, $M_{control}$ = 2.83, $t$ = -3.51, $p$ = 0.001). This indicated the experimental manipulation of ChatGPT fluency was successful. Besides, participants in the experimental group reported significantly higher levels of positive affect ($M_{experimental}$ = 3.31, $M_{control}$ = 2.89, $t$ = -3.98, $p < 0.001$). and GenAI distortion (subjective report) ($M_{experimental}$ = 2.57, $M_{control}$ = 2.30, $t$ = -2.23, $p$ = 0.020) than participants in the control group. Although the experimental group's posttest objective GenAI distortion scores were higher than those of the control group, they did not show a significant difference ($M_{experimental}$ = 1.30, $M_{control}$ = 1.19, $t$ = -0.62, $p$ = 0.535).

Table 4. Independent $t$-tests on variables.

| Variables | Experimental group ($N$ = 97) | Control group ($N$ = 78) | $t$ value | $p$-value |
|---|---|---|---|---|
|  | $M$ ($SD$) | $M$ ($SD$) |  |  |
| Pretest objective GenAI distortion | 0.48(0.82) | 0.28(0.62) | -1.86 | 0.065 |
| Pretest subjective GenAI distortion | 2.60(0.71) | 2.42(0.59) | -1.71 | 0.089 |
| Posttest objective GenAI distortion | 1.30(1.09) | 1.19(1.17) | -0.62 | 0.535 |
| Posttest subjective GenAI distortion | 2.57(0.83) | 2.30(0.65) | -2.23** | 0.020 |
| Current emotional state | 3.28(0.67) | 3.20(0.69) | -0.79 | 0.434 |
| Positive affect | 3.31(0.66) | 2.89(0.71) | -3.98*** | 0.000 |
| Manipulation check | 3.26(0.78) | 2.83(0.81) | -3.51*** | 0.001 |
| Task difficulty | 3.41(0.72) | 3.4(0.73) | -0.17 | 0.892 |

*Note.* $^*p < 0.01$, $^{***}p < 0.001$.

### 5.2.3. Tests of the mediating effect

We further examined the mediating effect of positive affect in the relation between manipulated GenAI fluency and GenAI distortion (Figure 1). We created a dummy variable as the independent variable. The control group was coded as 0, while the experimental group was coded as 1.

Firstly, a direct path from GenAI fluency to GenAI distortion was formulated. The results showed that manipulated GenAI fluency didn't influence the objective GenAI distortion after controlling for the covariables of sex and age ($\beta = 0.107$, $p = 0.535$). However, manipulated GenAI fluency positively influenced the subjective GenAI distortion after controlling for the covariables of sex and age ($\beta = 0.269$, $p = 0.020$).

Secondly, we used the SPSS PROCESS macro to test the mediation model (Model 4; Hayes, 2013). After controlling the current emotional state and the subjective distortion of the pre-test, the mediating effect of positive affect was significant between the manipulated GenAI fluency and subjective GenAI distortion ($\beta = 0.051$, 95% CI = [0.006, 0.121]) (see Figure 1). However, the mediating effect of positive affect was not significant between the manipulated GenAI fluency and objective GenAI distortion ($\beta = 0.039$, 95% CI = [-0.047, 0.136]).

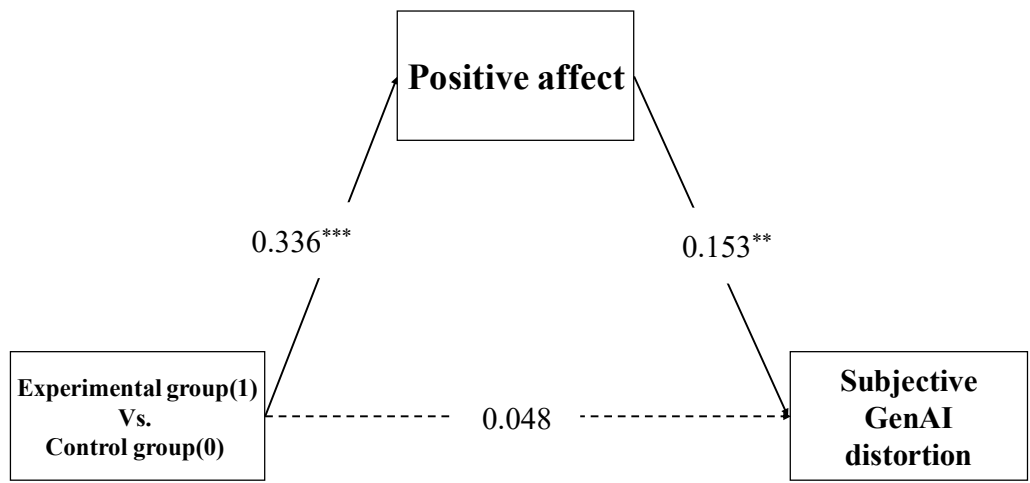

**Figure 1**. The mediation model

*Note.* Standardized estimates are shown. Dashed lines indicate nonsignificant paths. [**]$p < 0.01$, [***]$p < 0.001$.

**5.3. Discussion**

Study 3 investigated the underlying mediating process by experimentally manipulating the GenAI fluency, and subsequently measuring positive affect and GenAI distortion. Our findings provided additional support that positive affect significantly mediated the relationship between GenAI fluency and GenAI distortion. Specifically, students who perceived more GenAI fluency have a more positive affect, leading to a greater tendency towards GenAI distortion than students who perceived less GenAI fluency group.

**6. Overall Discussion**

To investigate the performance of college students in using GenAI, we introduce a new variable called GenAI distortion. Through qualitative research, cross-sectional surveys, and experimental studies, this study provides new empirical evidence for GenAI distortion, as well as the GenAI fluency and positive affect associated with GenAI distortion. The summary of the methods and results of the three studies is presented in Table 5. Notably, the study emphasizes how users' perception of GenAI fluency can enhance the degree of GenAI distortion through their positive affect. These findings offer new perspectives on guiding students to use GenAI scientifically, highlighting the intrinsic correlation of GenAI distortion within the context of GenAI.

Table 5. Summary of the methods and results of the studies.

| Study | Participants | Design | Procedures | Analyses | Findings |
|-------|-------------|--------|-----------|----------|----------|
| 1 | $N = 10$ | Semi-structured interview | ·Identified the manifestations of $Y$ <br> ·Explored key influencing factors ($X$, $M$) | Themes analysis | ·$Y$ manifests in a variety of ways <br> ·$X$, $M$ may influence $Y$ |

| 2 | $N = 999$ | Questionnaire survey | Measured $X, M, Y$ | ·Descriptive statistics ·Correlation analysis ·SEM analyses | Cross-sectional associations between $X, M, Y$ |
|---|---|---|---|---|---|
| 3 | $N = 175$ | Single-factor between-subject experiment | ·Manipulated $X$ ·Measured $M, Y$ | ·Independent sample $t$-tests ·The mediation model analyses in PROCESS | ·$X \rightarrow M$ ·$M \rightarrow Y$ |

## 6.1. The manifestation of GenAI distortion

The fluency of GenAI makes it convenient for users, prompting many college students to rely on GenAI to complete tasks and solve problems. However, through qualitative research, we further found that using GenAI is challenging because GenAI may distort scientific facts, potentially leading to biases and the spread of false information. The findings from our qualitative study lend support to the non-empirical conjectures previously made by researchers (Kidd & Birhane, 2023). For the first time, through interviews, we have discovered in a small sample that individuals may trust content produced by Generative AI systems, leading to distorted cognitions such as trust bias, misleading bias, and unconscious bias. This may further influence decision-making processes, potentially leading to poor judgments and decisions based on inaccurate information.

Additionally, based on qualitative material analysis, we establish a connection between GenAI fluency, positive affect, and GenAI distortion from a subject-object perspective. We found that GenAI typically exhibits certain characteristics that users perceive, such as anthropomorphic features, rapid user interaction, and the smooth delivery of seemingly reliable answers. These characteristics, especially the smooth experience, often evoke a positive affect in users, increasing their likelihood of trusting the content produced by GenAI. This finding aligns with previous evidence in the fields of conceptual fluency and perceptual fluency regarding perceptual smoothness (Alter & Oppenheimer, 2009), and extends this effect to the GenAI domain. Users may be consciously aware of these fluencies in GenAI, but potential biases in the rapid responses are often overlooked, which could potentially distort users' beliefs (Kidd & Birhane, 2023). These findings provide us with a prior foundation for our subsequent research to verify the quantitative relationships, including correlations and causal prediction.

## 6.2. Positive affect mediates the relation between GenAI fluency and GenAI distortion

Two studies have made valuable contributions to examining the mediation model by combining survey investigations with experimental manipulations of independent variables. The findings from one survey study and one experimental study provide consistent evidence that positive affect mediated the relationship between GenAI fluency and GenAI distortion. This finding extends the hedonic fluency model (Winkielman et al., 2003) and the affect as information theory (Schwarz, 2012) to the GenAI domain, indicating that GenAI fluency triggers users' positive affect, making them more likely to generate GenAI distortion under the catalysis of positive affect. Topolinski and Strack (2009) explained that when stimuli undergo processing fluently, the person doing the processing tends to display a subtle. This subtle facial activity leads to a positive affect triggered by free-floating fluency, which in turn generates intuitive feelings of correctness at the fringe of consciousness. In other words, the experience of cognitive ease (e.g., perceived fluency) leads to a vague pleasant feeling that emerges in awareness, which may subsequently produce biased cognitive judgments (Trent et al., 2013). GenAI distortion is a prerequisite for judgments, thus leading to the occurrence of distortion. Thus, students who perceive GenAI fluency influence the levels of GenAI distortion through the role of positive affect.

## 6.3. Limitations and future directions

This study has several potential limitations. Firstly, all collected data regarding positive affect rely on students' self-reports. While self-report data is commonly used to assess emotions (Schmukle et al., 2002), it is recommended that future research utilize computer-assisted emotion recognition systems, physiological measures (such as fMRI and EEG), as well as behavioral observation and coding (including facial expression recognition, language, intonation, and body posture) to enhance objective measurement metrics and improve the study's objectivity. Furthermore, this study only employed one reasonable paradigm to manipulate GenAI fluency. While this paradigm is widely regarded as effective (Flavián et al., 2017), recent research has identified multiple manipulations of perceptual fluency (e.g., visual search task, word game task, etc.) (Periáñez et al., 2021; Williams et al., 2020). Future research could utilize multiple paradigms to manipulate GenAI fluency, thereby enhancing the causal effectiveness of this study. Finally, this study was conducted within an Asian context. Considering the differences in attitudes towards GenAI across various countries and regions, it is essential for future research to conduct cross-cultural validation.

## 6.4. Methodological, theoretical, and practical implications

Regarding methodological implications, our study combines qualitative studies, cross-sectional surveys, and experimental manipulations to validate the relationship mechanism between GenAI fluency and GenAI distortion. The combination and design of different methods enhance the reliability of the research results.

Regarding theoretical implications, in the context of GenAI, we have introduced a new concept—GenAI distortion. Previous research on distortion has been more focused on mental health. This study, however, combines the characteristics of GenAI and its potential to distort individual beliefs and introduces the new concept of GenAI distortion. Besides, this study adopts the perspective of object-subject, revealing the mediating role of positive emotions (subject) between GenAI fluency (object) and GenAI distortion (subject). Additionally, our study extends the application of the hedonic fluency model and the affect of information theory on the domain of GenAI distortion. Existing research has mostly applied them to the general decision domain (Fang et al., 2007; Schwarz, 2012). However, with the emergence of GenAI characterized by fluency response (Kidd & Birhane, 2023) and high anthropomorphism (Alabed et al., 2023), it is uncertain whether these theories can adequately explain user belief distortion in the context of GenAI. Our study empirically addresses this issue.

Our research has practical value in promoting users to use GenAI more rationally and responsibly. This study found that GenAI fluency triggers positive emotional responses in users. However, this positive affect can lead to users relying on GenAI's answers, overlooking potential inaccuracies, thus distorting their beliefs. Based on these findings, we believe that combining measures such as education, technological improvements, and user guidance can help college students better address the GenAI distortion.

## 7. Conclusion

GenAI has attracted many users with its fluency and rapid responses. However, these characteristics can also lead to a potential distortion of user beliefs. To explore the impact of GenAI's fluency features on users' emotions and cognition, we selected university students as our study subjects and conducted three types of research—qualitative, correlational, and causal—based on the hedonic fluency model. Through qualitative analysis, we identified three main manifestations of GenAI distortion: trust bias, misleading bias, and unconscious bias. Additionally, both a cross-sectional study and an experimental manipulation approach demonstrated the role of positive affect in the correlation between GenAI fluency and GenAI distortion. Our findings indicate that users who perceived GenAI fluency increased their positive affect, which in turn facilitated the possibility of GenAI distortion among students. This study expands the theoretical boundaries of fluency in the GenAI domain and encourages users to be aware of and guard against potential errors in practical GenAI usage.